\documentstyle[12pt,epsfig]{article}
\def\3{\ss}                                                                                        
                                                                           
\renewcommand{\author}{ }  
\begin{document}
\vspace{1 cm}
%
%---------- Abbreviations -------------
%
\newcommand{\bec}       {\begin{center}}
\newcommand{\eec}       {\end{center}}
\newcommand{\qsq}       {\mbox{$Q^{2}$}}
\newcommand{\jpsi}      {\mbox{$J/\psi$}}
\newcommand{\psip}      {\mbox{$\psi(2S)$}}
\newcommand{\upsi}      {\mbox{$\Upsilon$}}
\newcommand{\Acce}	{\mbox{$\mathcal{A}$}}
\newcommand{\Lumi}	{\mbox{$\mathcal{L}$}}
\newcommand{\BR}	{\mbox{$\mathcal{B}$}}
\newcommand{\mv}        {\mbox{$m_{V}$}}
\newcommand{\wgp}       {\mbox{$W_{\gamma p}$}}
\newcommand{\zv}        {\mbox{$z_{\rm vert}$}}
\newcommand{\sleq}      {\raisebox{-.6ex}{${\textstyle\stackrel{<}{\sim}}$}}
\newcommand{\sgeq}      {\raisebox{-.6ex}{${\textstyle\stackrel{>}{\sim}}$}}
\newcommand{\Gev}       {\mbox{${\rm GeV}$}}
\newcommand{\Gevsq}     {\mbox{${\rm GeV}^2$}}
\newcommand{\x}         {\mbox{${\it x}$}}
\newcommand{\smallqsd}  {\mbox{${q^2}$}}
\newcommand{\ra}        {\mbox{$ \rightarrow $}}
\newcommand{\gamv}      {\mbox{$\Gamma_{V}$}}
%
% ---- commands from alessia -----
%
\newcommand{\sigmabrep}  {\mbox{$0.68\pm0.30(stat.)^{+0.14}_{-0.12}(syst.)$}}
\newcommand{\sigmabrept}  {\mbox{$0.68\pm0.30^{+0.14}_{-0.12}$}}
\newcommand{\sigmabrgp}  {\mbox{$13.3\pm 6.0(stat.)^{+2.7}_{-2.3}(syst.)$~}}
\newcommand{\sigmabrgpt}  {\mbox{$13.3\pm 6.0^{+2.7}_{-2.3}$}}
\newcommand{\sigmagpos}  {\mbox{$ 375\pm 170(stat.)^{+75}_{-64}  (syst.)$}}
\newcommand{\ratioupspsi}
{\mbox{$(4.8\pm 2.2(stat.)^{+0.7}_{-0.6}(syst.))\times 10^{-3}$}}
\newcommand{\ratioupspsit}
{\mbox{$(1.7\pm 0.8^{+0.2}_{-0.2})\times 10^{-4}$}}
\newcommand{\ratioupspsito}
{\mbox{$1.7\pm 0.8^{+0.2}_{-0.2}$}}
%
% ---- commands from paul -----
%
\parskip 2mm plus 1mm minus 1mm
\def\ctr#1{{\it #1}\\\vspace{10pt}}
\def\si{{\rm si}}
\def\Si{{\rm Si}}
\def\Ci{{\rm Ci}}
\def\px{p_{_{x}}}
\def\py{p_{_{y}}}
\def\pz{p_{_{z}}}
\def\yjb{y_{_{JB}}}
\def\xjb{x_{_{JB}}}
\def\qjb{\qsq_{_{JB}}}
\def\gap{\hspace{0.5cm}}
\renewcommand{\thefootnote}{\arabic{footnote}}
%
% --------  Title, Date and Authors
%
\begin{flushright}{\large DESY-98-089}  \end{flushright}
\vspace{1cm}
\bec{\bf
%     DRAFT DRAFT DRAFT DRAFT DRAFT DRAFT DRAFT
     }\eec
\vspace{1cm}
\bec{\Large\bf 
    Measurement of Elastic $\Upsilon$ Photoproduction at HERA
     }\eec
\vspace{1cm}
\vspace{1.0cm}
\bec{\bf 
    ZEUS Collaboration
    }\eec
\vspace{3cm}
%
% -------- Abstract -------------------
%
\begin{abstract}
\centerline{\vbox{\hsize13.7cm
\noindent
The photoproduction reaction
$\gamma p \rightarrow  ~\mu^+ \mu^- p$ has been
studied in $ep$ interactions
using the ZEUS detector at HERA. The data sample corresponds to 
an integrated luminosity of $43.2\ {\rm pb}^{-1}$. 
The $\Upsilon$ meson has been observed in photoproduction
for the first time.
The sum of the products of the elastic  $\Upsilon(1S), \Upsilon(2S), 
\Upsilon(3S)$ photoproduction cross sections with their respective 
branching ratios is determined to be \sigmabrgp pb at a mean
photon-proton centre of mass energy of $120$ GeV.
The cross section is above the prediction of a perturbative 
QCD model.
}}
\end{abstract}
%
%--------  Reset page counter and goto next page
%
\setcounter{page}{0}
\thispagestyle{empty}
\pagenumbering{Roman} 
\newpage
%===================================================================                               
%                                                                                                  
%  MEMBER NAME  AUTH63 (ZEUS)     M  TEX                                                           
%                                                                                                  
%  JH.: transformed to a format, which is suited as input for                                      
%       CONVERT, which automatically creates author-indices                                        
%                                                                                                  
%  Don't remove lines starting with a percent sign %,                                              
%  CONVERT may need them urgently !                                                                
%                                                                                                  
%=====================================================================                            
\begin{center}                                                                                     
{                      \Large  The ZEUS Collaboration              }                               
\end{center}                                                                                       
  J.~Breitweg,                                                                                     
  M.~Derrick,                                                                                      
  D.~Krakauer,                                                                                     
  S.~Magill,                                                                                       
  D.~Mikunas,                                                                                      
  B.~Musgrave,                                                                                     
  J.~Repond,                                                                                       
  R.~Stanek,                                                                                       
  R.L.~Talaga,                                                                                     
  R.~Yoshida,                                                                                      
  H.~Zhang  \\                                                                                     
 {\it Argonne National Laboratory, Argonne, IL, USA}~$^{p}$
% ...
\newpage                                                                                           
% ...
$^{  19}$ now at ATLAS Collaboration, Univ. of Munich \\                                           
% ...
$^{  30}$ partially supported by the Foundation for German-Russian Collaboration                   
DFG-RFBR \\ \hspace*{3.5mm} (grant no. 436 RUS 113/248/3 and no. 436 RUS 113/248/2)\\              
% ...
                                                           %                                       
                                                           %                                       
% \par         % if index listing & table fit to 1 page, put gap here                              
\newpage   % alternatively: go to newpage, if page is too small                                    
                                                           %                                       
% \institute_references_start    % do not touch or move this line !                                
                                                           %                                       
\begin{tabular}[h]{rp{14cm}}                                                                       
$^{a}$ &  supported by the Natural Sciences and Engineering Research                               
          Council of Canada (NSERC)  \\                                                            
$^{b}$ &  supported by the FCAR of Qu\'ebec, Canada  \\                                            
$^{c}$ &  supported by the German Federal Ministry for Education and                               
          Science, Research and Technology (BMBF), under contract                                  
          numbers 057BN19P, 057FR19P, 057HH19P, 057HH29P \\                                        
$^{d}$ &  supported by the MINERVA Gesellschaft f\"ur Forschung GmbH,                              
          the German Israeli Foundation, the U.S.-Israel Binational                                
          Science Foundation, and by the Israel Ministry of Science \\                             
$^{e}$ &  supported by the German-Israeli Foundation, the Israel Science                           
          Foundation, the U.S.-Israel Binational Science Foundation, and by                        
          the Israel Ministry of Science \\                                                        
$^{f}$ &  supported by the Italian National Institute for Nuclear Physics                          
          (INFN) \\                                                                                
$^{g}$ &  supported by the Japanese Ministry of Education, Science and                             
          Culture (the Monbusho) and its grants for Scientific Research \\                         
$^{h}$ &  supported by the Korean Ministry of Education and Korea Science                          
          and Engineering Foundation  \\                                                           
$^{i}$ &  supported by the Netherlands Foundation for Research on                                  
          Matter (FOM) \\                                                                          
$^{j}$ &  supported by the Polish State Committee for Scientific                                   
          Research, grant No.~115/E-343/SPUB/P03/002/97, 2P03B10512,                               
          2P03B10612, 2P03B14212, 2P03B10412 \\                                                    
$^{k}$ &  supported by the Polish State Committee for Scientific                                   
          Research (grant No. 2P03B08614) and Foundation for                                       
          Polish-German Collaboration  \\                                                          
$^{l}$ &  partially supported by the German Federal Ministry for                                   
          Education and Science, Research and Technology (BMBF)  \\                                
$^{m}$ &  supported by the Fund for Fundamental Research of Russian Ministry                       
          for Science and Edu\-cation and by the German Federal Ministry for                       
          Education and Science, Research and Technology (BMBF) \\                                 
$^{n}$ &  supported by the Spanish Ministry of Education                                           
          and Science through funds provided by CICYT \\                                           
$^{o}$ &  supported by the Particle Physics and                                                    
          Astronomy Research Council \\                                                            
$^{p}$ &  supported by the US Department of Energy \\                                              
$^{q}$ &  supported by the US National Science Foundation \\                                       
\end{tabular}                                                                                      
                                                           %                                       
% \institute_references_end     % do not touch or move this line !                                 
%
%--------  Reset page counter and goto next page
%                                                           %                                       
\newpage
\setcounter{page}{1}
\pagenumbering{arabic}                                                                                   
%
%--------  Introduction and Motivation
%
\section{\bf Introduction}
\label{Sect:Intro}

%Elastic processes are commonly described through phenomenological 
%non-perturbative methods incorporating soft pomeron exchange.
Perturbative QCD can be applied to $ep$ scattering to calculate 
the amplitude for elastic production of heavy vector mesons.
Previous HERA results on $J/\psi$ meson production
at $Q^2\simeq 0$ and for $0.25<Q^2<40$~GeV$^2$ \cite{ZeusJPsi94, H1JPsi94}
have shown that the rise of the cross section with $W$ as well as the 
dependence of the cross section with $Q^2$  can be described by
perturbative QCD models \cite{Brodsky,Ryskin,Frankfurt}.
Only upper limits of $\sigma \cdot \BR$  exist in the literature
for inclusive $\Upsilon$ production in lepton-hadron collisions \cite{i2}.

In this paper we extend the study of elastic photoproduction 
of vector mesons at HERA to $\Upsilon$ mesons.
The measurement is made with the ZEUS detector using a data sample which
corresponds to an integrated $ep$ luminosity of $43.2$ pb$^{-1}$.
The improved luminosity allows the study of the reaction 
$\gamma p \rightarrow \mu^+\mu^- p$
for $\mu^+\mu^-$ invariant masses beyond the $\Upsilon$ mass region.
A first measurement of 
$\sigma_{\gamma p \rightarrow \Upsilon p} \cdot \BR$ 
in photoproduction is presented 
in the kinematic range of the photon-proton centre of mass energy 
$80<W<160$ GeV, where $\BR$ is the $\Upsilon$ branching ratio to muons.
The $\Upsilon(1S)$, $\Upsilon(2S)$ and $\Upsilon(3S)$ resonances 
are not resolved in this measurement.
Under the assumption that the production ratios of 
$\Upsilon(1S)$, $\Upsilon(2S)$ and $\Upsilon(3S)$
are the same as those measured in hadron-hadron collisions \cite{CDFups95, i1}, 
we determine the $\Upsilon(1S)$ and the ratio of the $\Upsilon(1S)$
to $J/\psi$ photoproduction cross sections
and compare them to a pQCD inspired model \cite{Frankfurt}.

%%%%%%%%%%%%%%%%%%%%%%%%
% Experimental Setup   %
%%%%%%%%%%%%%%%%%%%%%%%%

\section{\bf Experimental Conditions}
\label{Sect:ZEUS}

In the years 1995-97 HERA collided positrons of 27.5 GeV with  protons 
of 820 GeV, corresponding to a centre of mass energy 
$\sqrt{s}=300$ GeV.
A description of the ZEUS detector can be found in references~\cite
{sigtot_photoprod,Detector}. The primary components used in this analysis
were the central tracking detector, the  
uranium-scintillator calorimeter, the muon chambers and the proton remnant
tagger.
The central tracking detector (CTD) \cite{CTD} operates in a $1.43\,{\rm T}$
solenoidal magnetic field. 
It is a drift chamber consisting 
of 72 cylindrical layers, organized into 9 superlayers.
It was used to identify the vertex and to
measure the  momenta and directions of the muons. The
transverse momentum resolution is $\sigma (p_t)/p_t=\left[ 0.005p_t
\right] \oplus 0.016$, with $p_t$ in GeV, for full length tracks. 
The calorimeter (CAL) ~\cite{CAL}
is hermetic.
%and consists of 5918~cells each read out by two photomultiplier tubes. Under 
Under test beam conditions, it has
energy resolutions of 18\%/$\sqrt{E}$ for electrons and 35\%/$\sqrt{E}$ for
hadrons. The time resolution is below $1\,{\rm ns}$ for energy
deposits greater than 4.5 GeV. The CAL was used to reject cosmic rays and 
beam halo background by timing and to identify minimum ionizing particles 
(m.i.p.).
It is surrounded by a magnetized iron yoke with a field
of $1.4\,{\rm T}$ produced by conventional warm coils. 
The muon system  (MUO) consists of tracking detectors (forward, barrel and
rear muon chambers: FMU~\cite{Detector}, B/RMU ~\cite{BRMU})
placed inside and outside the yoke covering 
the polar angles from $10^{\circ}$ to $171^{\circ}$.
They were used for the  trigger and to tag the muons 
by matching segments in the MUO
chambers with tracks in the CTD and m.i.p.'s in the CAL.
The proton remnant tagger (PRT) consists of two stations of scintillation
counters surrounding the beamline at $Z=5~\rm m$ and $Z=24~\rm m$ 
\footnote{%
The right-handed ZEUS coordinate system is centred on the nominal
interaction point ($Z=0$) and defined with the $Z$ axis pointing in the
proton beam direction, and the horizontal $X$ axis pointing towards the
centre of HERA.}, 
and tags particles  in the forward proton direction in an angular 
range 6 to 26 mrad and 1.5 to 8 mrad, respectively. 
It was used to estimate proton dissociative background.
The luminosity was measured to a precision of $1.5\%$ from the rate
of energetic bremsstrahlung photons produced in the process 
$e^+p\rightarrow e^+ \gamma p$.
%%%%%%%%%%%%%%%%%%%%%
% Kinematics        %
%%%%%%%%%%%%%%%%%%%%%

\section{\bf Kinematics}
\label{Sect:Kinem}

Figure \ref{fig_Feynman}a shows a schematic diagram for the reaction
\begin{equation}
 e^{+}(k)~p(p) ~\rightarrow ~e^{+}(k')~V(v)~p(p'),
\label{eq-V}
\end{equation}
where the symbol in parenthesis denotes the four-momentum of the 
corresponding particle, and $V$ indicates a \jpsi ~or an $\Upsilon$. 

The kinematics of the inclusive scattering of 
unpolarised positrons and protons are
described by the positron-proton centre of mass energy squared ($s$)
and any two of the following variables:
\begin{itemize}
    \item $Q^2=-q^2=-(k-k')^2$, the negative four-momentum squared of the 
         exchanged photon; 
    \item $y=(q\cdot p)/(k\cdot p)$, the fraction of the positron energy 
         transferred to the hadronic final state in the rest frame of the 
         initial state proton;
    \item $W^2 = (q+p)^2= -Q^2+2y(k\cdot p)+M^2_p$, the
         centre of mass energy squared of the photon-proton system,
         where $M_p$ is the proton mass.
\end{itemize}

For a complete description of the exclusive reaction 
$e^{+}p \rightarrow e^{+} ~V~ p$
($V \rightarrow \mu^{+}\mu^{-}$) the following additional variables are
required: 
\begin{itemize}
    \item $M_V$, the invariant mass of the $\mu^+\mu^-$ pair;
    \item $t = (p-p')^2$, the four-momentum transfer squared at the proton 
         vertex;
    \item $\Phi$, the angle between the vector meson production plane 
         and the positron scattering plane in the photon-proton centre
	 of mass frame;
    \item $\theta_h$ and $\phi_h$, the polar and azimuthal angles  of
         the positively charged decay lepton in the $V$ helicity frame.
\end{itemize}

In this analysis, photoproduction events are selected by requesting
a $ \mu^+\mu^-$~ pair from the interaction point and nothing else 
in either the CTD or the CAL.
For the selected events the $Q^2$ value ranges from the 
kinematic minimum $Q^2_{min} = M^2_e y^2/(1-y) \approx 10^{-9}~\rm{GeV^2}$, 
where $M_e$ is the electron mass, to the value at which the scattered positron
starts to be observed in the uranium calorimeter 
$Q^2_{max} \approx 1 \ \rm{GeV^2}$, with a median $Q^2$ of
approximately $5 \times 10^{-5} {\rm~GeV^2}$. 
Since the typical $Q^2$ is small, the photon-proton centre of mass energy can
be expressed as
\begin{equation}
    W^2 \approx 4 E_p E_e y \approx 2E_p (E - p_{Z})_V ,
\label{Eq:W2Def}
\end{equation}
where $E_p$ is the laboratory energy of the
incoming proton and $(E - p_{Z})_V$ 
is the difference between the
energy and the longitudinal momentum of the vector meson, $V$,
as determined from the CTD tracks assuming the muon mass.
 $\Phi$ is not measurable since the scattered positron is not detected.

%%%%%%%%%%%%%%%%%%%
% Event Selection %
%%%%%%%%%%%%%%%%%%%

\section{\bf Event Selection}
\label{Sect:evsel}

Elastic $\mu^+\mu^-$ events were selected using dedicated triggers.
Trigger cuts are superseded by the following offline cuts:

%The FMU and B/RMU trigger chains are largely independent and therefore
%allow a study of the trigger efficiency. The study yields efficiencies, for
%muons with momenta $> 2$ GeV, of approximately $60$\%~ over all
%angular coverage.

\begin{itemize}
    \item CAL timing and reconstructed interaction vertex consistent with 
          the nominal $ep$ interaction to reject non $e^+p$ background;
    \item two oppositely charged tracks from the vertex,
          at least one of which matches a segment in the B/RMU chambers or a hit in
          the FMU chambers,
          and no other track in the CTD;
    \item at least 3 CTD superlayers per track, limiting the polar angular 
          region from $\simeq 17^{\circ}$ to $\simeq 163^{\circ}$;
    \item acollinearity of the tracks $\Omega<174^{\circ}$ where 
          $\Omega$ is the angle between the two tracks, to reject 
          cosmic ray events;
    \item invariant mass of the two CTD tracks, treated as muons, 
          larger than 2 GeV;
    \item CAL energy associated to each track consistent with 
          the energy deposit of a m.i.p.;
%    \item the two tracks are requested to match two energy deposits in CAL 
%          compatible with a m.i.p. signal, where a m.i.p. is defined as
%          energy between 0.8 and 5 GeV 
%          with a ratio of at least 0.8 between the energy
%	  in the hadronic and in the electromagnetic sections;
    \item no CAL cell (apart from those associated with a $\mu$) 
          with energy greater
          than 150 MeV, well above the CAL uranium noise level.
\end{itemize}

The events were then selected in the kinematical range of $W$ 
between 80 and 160 GeV, corresponding to an acceptance of 
$\approx 40 \%$ for the \upsi; 
the acceptance for the \jpsi~falls smoothly from
$ 38\% ~\rm to ~ 10 \%$ with increasing $W$ over this $W$ range.     

%The effect of varying the cuts is discussed in Section \ref{subs:syst}.

%%%%%%%%%%%%%%%
% Montecarlo  %
%%%%%%%%%%%%%%%
\section{\bf Monte Carlo Simulation }
\label{Sect:MC}

\begin{figure}
\begin{center}
\epsfig{file=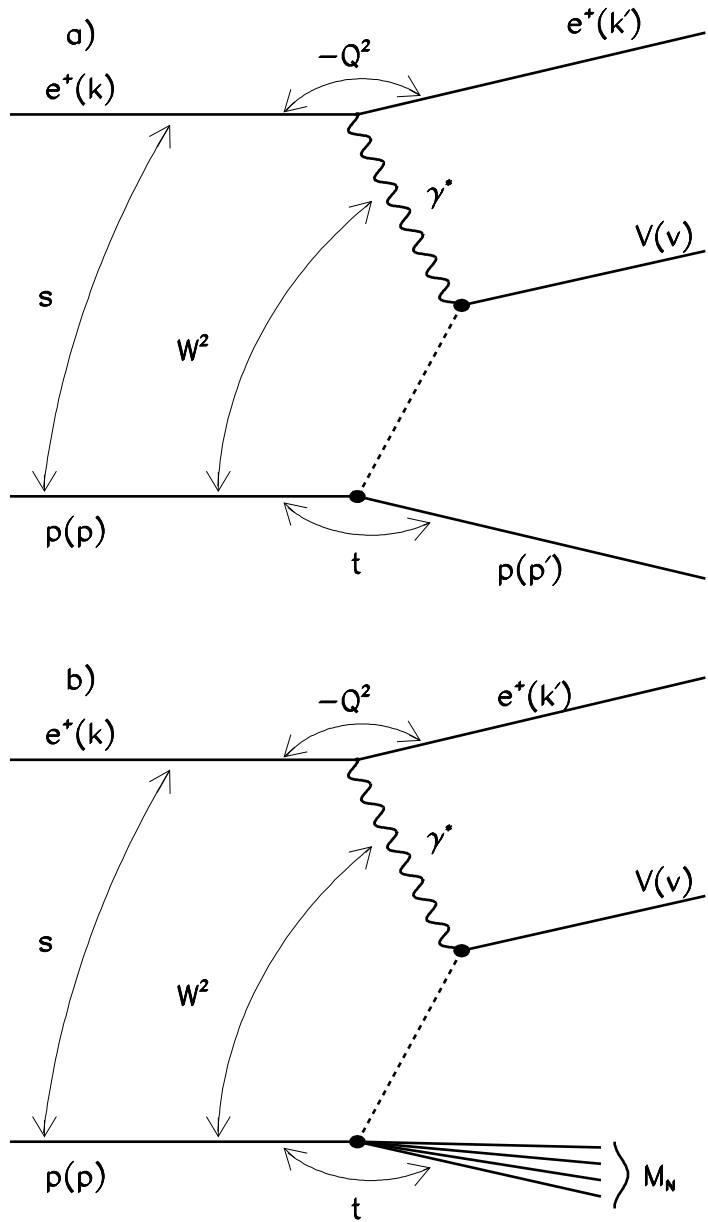,width=10cm}
%      height=15cm,%
%       width=15cm,%
%       bbllx=0pt,%
%       bblly=0pt,%
%       bburx=567pt,%
%       bbury=567pt,%
%     rheight=15cm,%
%      rwidth=15cm,%
%        clip=}
\caption{
         Schematic diagrams for diffractive vector meson ($V$)~electroproduction.
         (a) Elastic $V$~production. 
         (b) Proton dissociative $V$~production where the proton 
         dissociates into a hadronic system of invariant mass $M_N$.
	 The dotted line represents a colorless exchange between the virtual
         photon and the proton.
         }
\label{fig_Feynman}
\end{center}
\end{figure}

To compute the acceptance, the reaction~\ref{eq-V}
(Figure \ref{fig_Feynman}a)
was simulated using the DIPSI generator \cite{DIPSI}.
DIPSI is based on a model developed by Ryskin \cite{Ryskin} in
which the exchanged photon fluctuates into a
$q\bar{q}$ pair which interacts with a gluon ladder emitted by
the incident proton. The parameters of the model are the strong
coupling constant $\alpha_s$ (assumed fixed), 
the two-gluon form factor and the gluon momentum density of the proton. 
The cross section dependence on $W$ and $t$ is fixed by these parameters. 
%The effect of the variation of the parameters 
%on the acceptance is discussed in Section \ref{subs:syst}.
 
Proton dissociative vector meson production, $e^+ p \rightarrow e^+ V N$
(Figure \ref{fig_Feynman}b), 
was simulated using the EPSOFT  \cite{Kasprzak}
and the DIFFVM \cite{diffvm} generators.
Both are based on the assumption that the diffractive cross section at large 
$M_N$ is of the form           
$d^2\sigma / d|t| dM_N^2\propto {e^{-b |t|}}/{M_N^{\beta}}$
where $M_N$ is the mass of the dissociative system;
we have used $b=1$ GeV$^{-2}$ and $\beta=2.2$.
The simulation of the dissociative system includes
a parametrisation of the spectrum in the resonance region 
which differs for the two generators.

The background Bethe-Heitler process, in which a lepton pair 
is produced by the
fusion of a photon radiated by the positron with a photon radiated
by the proton, was simulated using the LPAIR 
generator \cite{Vermaseren}. Both elastic and proton dissociative
events were generated.

All Monte Carlo events were passed through a 
simulation of the ZEUS detector and trigger
based on the GEANT program \cite{geant}
and analysed with the same
reconstruction and offline selection procedures as the data.
The overall acceptance in a selected kinematic range was obtained 
as the ratio of the number of Monte Carlo events 
passing the cuts to the number of events generated in the same  range.
The acceptance, calculated in this manner, accounts for the geometric
acceptance, for the detector, trigger and reconstruction efficiencies, and 
for the detector resolution.

%%%%%%%%%%%%%%%%%%%%%%
% Analysis           %
%%%%%%%%%%%%%%%%%%%%%%

\section{\bf Analysis}
\label{Sect:Analysis}

The measured $\mu^+\mu^-$ mass distribution, for the sample of  events  
obtained by the selection described in Section \ref{Sect:evsel}, is
shown in Figure \ref{fig_mass}. 
The signals of the \jpsi,~\psip~ and $\Upsilon$'s (unresolved) are apparent. 
The continuum is well described by the Bethe-Heitler process apart from an
enhancement in the region of $6$ GeV, which is consistent with being 
a fluctuation.

\subsection{\bf Extraction of the Signal}
\label{subs:signal}

\begin{figure}
\begin{center}
\epsfig{file=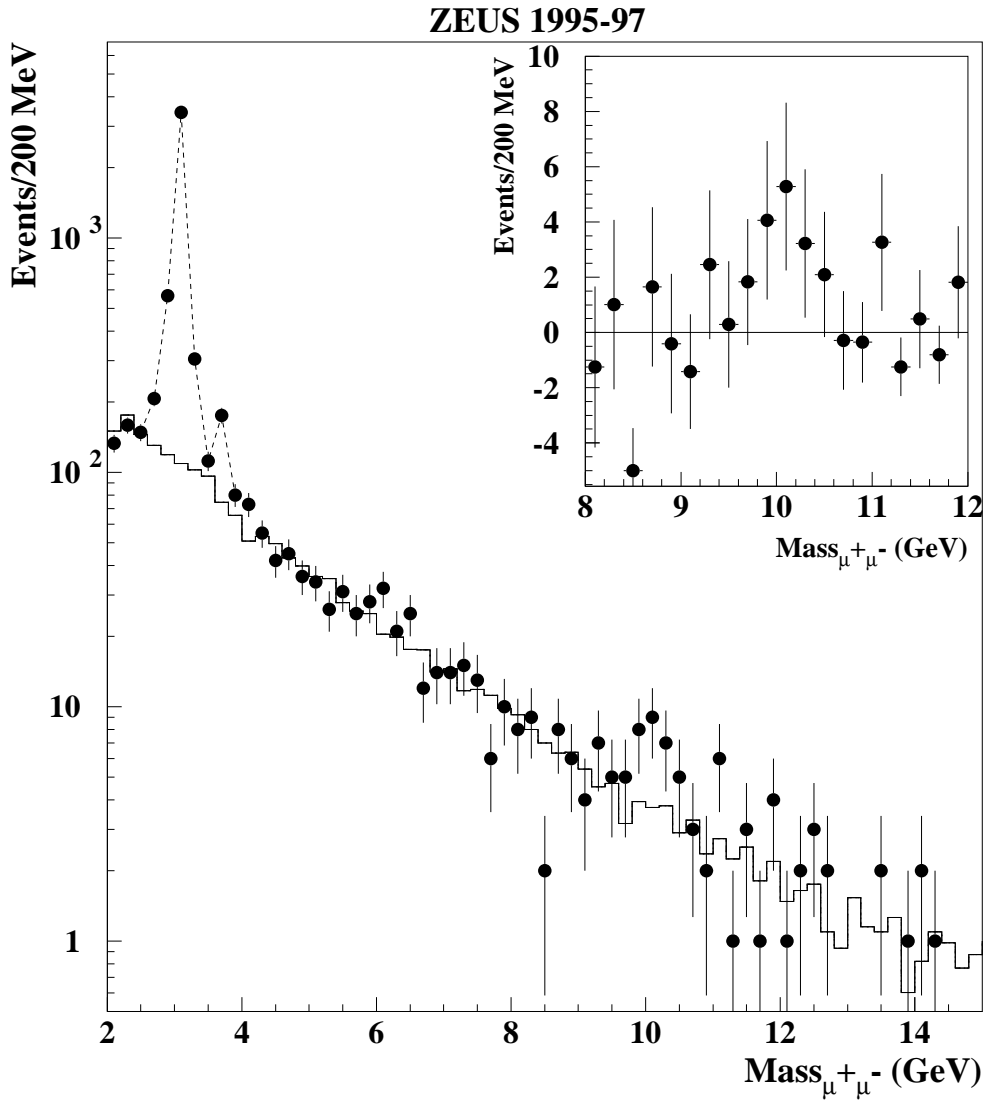,width=14cm}
\caption{
         Mass distribution of $ \mu^+ \mu^-$ pairs. The histogram
         represents the simulated Bethe-Heitler background. 
         Points in the \jpsi~ region are connected
         by a dotted line to guide the eye. The insert shows the signal
	 remaining in the \upsi~region after subtraction of the
	 non-resonant background.
         }
\label{fig_mass}
\end{center}
\end{figure}

The elastic Bethe-Heitler process can be calculated to good precision
since it is a pure QED reaction. There is, however, some 
arbitrariness in the parametrisation of the proton dissociative component. 
Comparing the data with the LPAIR simulation
in the mass interval 4 to 8 GeV,
where we expect the Bethe-Heitler process to dominate,  
we find that the  simulation underestimates the energy
deposited in the forward region of CAL due to proton dissociation.
This leads to an overestimate of the Bethe-Heitler events which 
pass the selection cuts 
if both the elastic and the proton dissociative processes 
are normalised to luminosity by their calculated
cross sections \cite{Vermaseren}.
However, since the shape of the $\mu^+\mu^-$ mass distribution predicted by LPAIR is the 
same for both processes, to evaluate this background we have
normalised Monte Carlo and data in a mass window not containing 
resonances (4.2  to 8.4 GeV). 
This corresponds to adding a $10$\% contribution to the
elastic Bethe-Heitler distribution normalised to the luminosity.
In Figure \ref{fig_mass} the data are compared with 
this renormalised Bethe-Heitler distribution.
The spectrum outside the resonance regions is
well reproduced and the distribution has been used to subtract the
background under the resonances.

The limited statistics and the $\mu^+\mu^-$ mass resolution of 0.3 GeV 
in the $\Upsilon$ region do not allow to distinguish between the
$\Upsilon(1S)$, $\Upsilon(2S)$ and $\Upsilon(3S)$ states.
The mass window 8.9 to 10.9 GeV
(i.e. from twice the resolution below the \upsi(1S) nominal mass, $9.46$ GeV,
to twice the resolution above the \upsi(3S) mass, $10.36$ GeV \cite{PDG}) 
was chosen to count the \upsi~events. 
In this region there are 57 events while 39.9 are estimated to be background.
The mass spectrum in the $\Upsilon$ region after background subtraction is shown 
in the insert of Figure \ref{fig_mass}; 
the mean mass and the r.m.s. for the background corrected
signal in the window are $9.9\pm0.2$ GeV and 0.47 GeV, respectively.
The simulation of the detector response to a mixture of $\Upsilon$ states  
\cite{CDFups95} yields a mean mass of $9.7$ GeV and an r.m.s. of $0.42$ GeV.
The r.m.s. is smaller since background was not considered.
% the mean value and the 
% %root mean square
% r.m.s.
% of the mass in the used window are $9.9\pm 0.2$ GeV and $0.47\pm 0.15$ GeV
% respectively. 
% The simulation of the detector response to a mixture of $\Upsilon$ states
% from \cite{CDFups95} yields a mean mass of 
% $9.7\pm 0.1$ GeV and a r.m.s. of $0.42 \pm 0.08$ GeV, in agreement 
% with the data.
%A simulation obtained using the same statistical significance as the data, 
%the DIPSI Monte Carlo mass resolution and the mixture
%of $\Upsilon$ states from \cite{CDFups95}, gives the corresponding 
%average values of $9.7\pm 0.1$ GeV and of $0.42 \pm 0.08$ GeV.
A total of 4257 events are counted
in the $J/\psi$ mass window from 2.8 to 3.35 GeV while 306 are expected 
from the Bethe-Heitler process.

The amount of proton dissociative resonant background remaining in the sample
after the cuts described in Section \ref{Sect:evsel}
is estimated using the number of events tagged by the PRT and the EPSOFT 
Monte Carlo to correct for the detector acceptance.
In the $J/\psi$ mass region the fraction of dissociative events is determined
to be $25\pm2(stat.)$\% . 
For $\Upsilon$ production, the same fraction of proton 
dissociative events has been assumed. 
Within the limited statistics, this assumption is consistent with 
the number of events in the $\Upsilon$ mass region tagged by the PRT.
 
\subsection{\bf Systematic Uncertainties}
\label{subs:syst}

A study of the systematic uncertainties
on the measurements has been performed. 
The systematic uncertainties, listed below, have been
divided into two classes, those common to the $\Upsilon$ and $J/\psi$ analysis,
which cancel in the measurement of the ratio, and those remaining.
\begin{itemize}
 \item Common systematic uncertainties:
 \begin{itemize}
  \item uncertainty on the CTD first level trigger efficiency, $\pm5$\%;
  \item uncertainty on the muon chamber and muon trigger efficiency,
  $\pm10$\%;
  \item uncertainty on the proton dissociation background
        estimated using DIFFVM, ($+9,-2$)\%.
 \end{itemize}
 \item  Systematic uncertainties specific to each meson:
 \begin{itemize}
 \item using a different mass region for normalisation of background contribution, 
        from 4 to 6 GeV, $+10$\%;
 \item varying the $\Upsilon$ mass window by $\pm 300 \rm$ MeV, $\pm 10$\%;
 \item varying the $\cos \theta_h$ distribution used in the Monte Carlo from
       that required by s-channel helicity conservation
        ($1+\rm \cos^{2} \theta_{\it h}$) to a flat distribution 
       for the $\Upsilon$ and between the limits allowed by the existing
       measurements for the $J/\psi$,  $-8\%$ for the $\Upsilon$, 
       $-4\%$ for the $J/\psi$;
 \item the uncertainties on the \jpsi~and~\upsi~muonic branching ratios,
 $\pm 3$\%.
 \end{itemize}

\end{itemize}
All of these uncertainties are added in quadrature yielding an overall
systematic error on $\sigma \cdot \BR$ of $(+20,-17)\%$. 
The systematics which affect the 
$\sigma_{\gamma p\rightarrow \Upsilon p}/ \sigma_{\gamma p\rightarrow J/\psi p}$ 
ratio are only those included in the second 
group which give contributions of $(+14,-13)\%$. 

Other possible sources of systematic uncertainty 
(variation of the energy thresholds
used to select the elastic events, removal of the m.i.p. requirement on the
energy associated with the tracks, variation of the \jpsi~mass window,
uncertainty in the luminosity determination, variation of the DIPSI parameters,
use of an unbinned fit to extract the $\Upsilon$ signal)
give negligible contributions. 

\section{\bf Results and comparison with QCD predictions}
\label{Sect:Results}

We have calculated $\sigma \cdot \BR$ as
\begin{equation}
\sigma_{e^{+} p \rightarrow e^{+} V p}\cdot \BR  = 
\frac{N_{evt} ~(1 - ~\cal{F} ~)} {\cal{A} ~\cal{L} },
\end{equation}
where $N_{evt}$ denotes the background subtracted number of ~$V$ signal events,
$\cal{F}$ is the estimated fraction of proton dissociative events,
$\cal{A}$ the acceptance, $\cal{L}$ the integrated luminosity  and 
$\cal{B}$ the muonic branching fraction.
Photoproduction cross sections were determined by dividing 
the electroproduction cross sections by the photon flux,
calculated according to reference \cite{Flux}. 

The production cross sections 
multiplied by the muonic branching ratios for the unresolved
$\Upsilon(1S)$, $\Upsilon(2S)$ and $\Upsilon(3S)$ are
summarized in Table \ref{Tab:xsect}.  
$\sigma_{\gamma p \rightarrow \Upsilon p} \cdot \BR$ is determined to be
\sigmabrgp pb at a mean photon-proton centre of mass energy
$\langle \rm W \rangle = 120$ GeV. 
The first error is statistical and the second is
the systematic uncertainty.
In Table \ref{Tab:xsect} we report also the ratio 
$\sigma_{\gamma p\rightarrow\Upsilon p}\cdot \BR$ summed on the $\Upsilon$
states over $\sigma_{\gamma p\rightarrow J/\psi p}$ in the $W$ range
between 80 and 160 GeV.

The QCD-based model of reference \cite{Frankfurt} 
gives predictions 
for the hard diffractive photo- and electroproduction
of heavy vector meson $J/\psi$ and $\Upsilon(1S)$
within the leading logarithmic approximation.
%$\alpha_s \ln \frac{Q^2}{\Lambda^2_{QCD}}$ QCD approximation.
To compare the data with the predictions, the experimental cross section
for $\Upsilon(1S)$ has to be derived. 
As our data does not allow the relative fractions of 
$\Upsilon(1S)$, $\Upsilon(2S)$ and $\Upsilon(3S)$ to be determined,
we have to make assumptions on their relative production rates.
Assuming that the cross sections times branching ratios are 
the same as measured by CDF \cite{CDFups95}, 
$\sigma\cdot\BR(\Upsilon(2S)) / \sigma\cdot\BR(\Upsilon(1S))
=0.281 \pm 0.030(stat.)\pm 0.038(sys)$ and
$\sigma\cdot\BR(\Upsilon(3S)) / \sigma\cdot\BR(\Upsilon(1S))
=0.155 \pm 0.024(stat)\pm 0.021(sys)$,
the $\Upsilon(1S)$ accounts for $70$\% of the signal.
Using the muonic branching ratio 
$\BR (\Upsilon (1S)\rightarrow \mu^+\mu^-)=(2.48\pm0.007)\%$ \cite{PDG}, 
we derive 
\begin{center}
$\sigma_{\gamma p\rightarrow \Upsilon(1S)~ p}= $ \sigmagpos~ pb  at $W=120$ GeV
\end{center}
\noindent and the ratio 
\begin{center}
$ \sigma_{\gamma p \rightarrow \Upsilon(1S)~ p}/
\sigma_{\gamma p \rightarrow J/\psi~ p} =$ \ratioupspsi.
\end{center}

\begin{figure}
\begin{center}
\epsfig{file=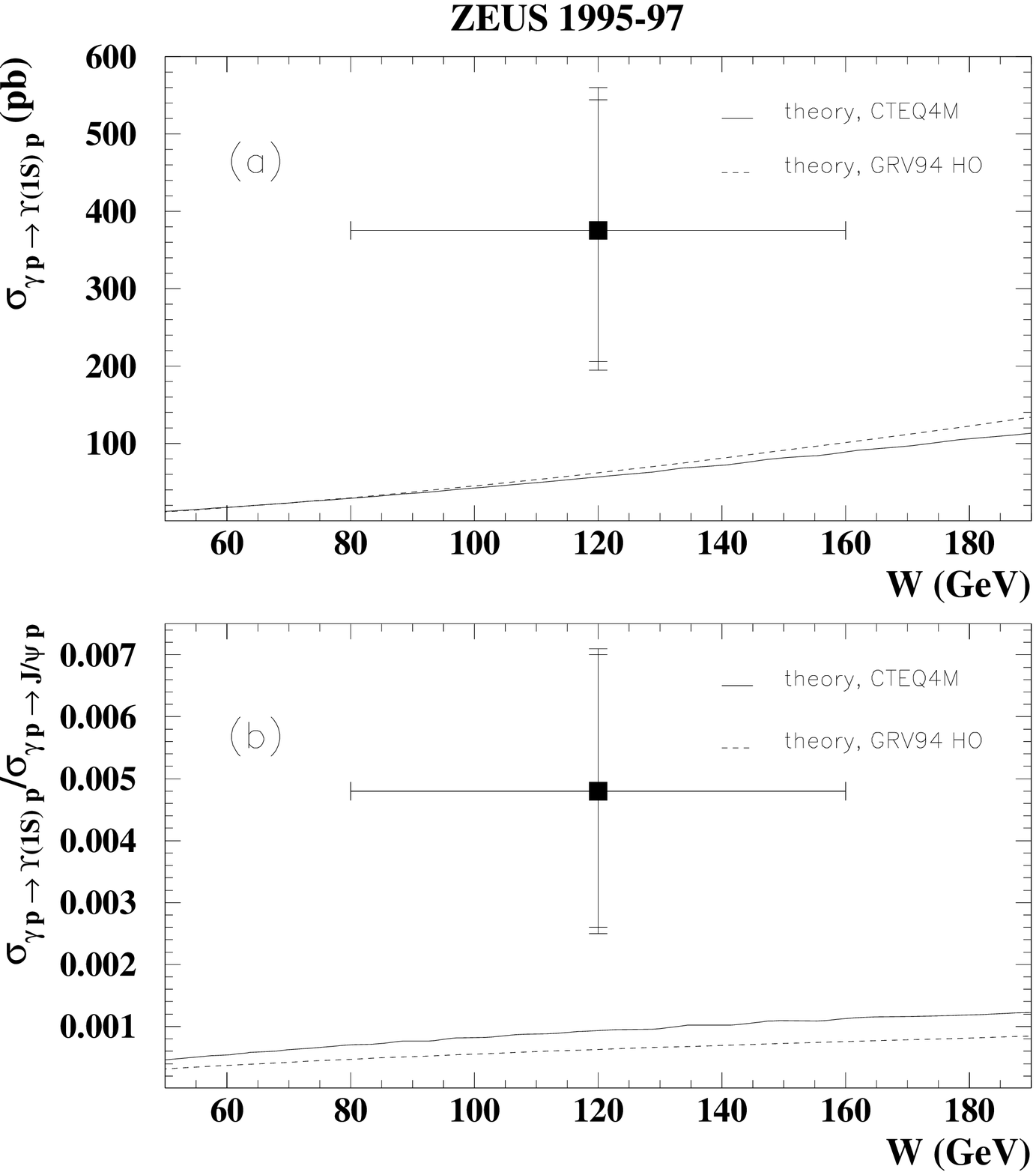,width=14cm}
\caption{
         (a)~ $\sigma_{\gamma p \rightarrow \Upsilon(1S)~ p} $ 
         and ~(b)~the
         ratio $ \sigma_{\gamma p \rightarrow \Upsilon(1S)~ p}/
	         \sigma_{\gamma p \rightarrow J/\Psi~ p} $
         as a function of $W$, the 
         centre of mass energy  of the photon-proton system. Data 
         (full squares) are compared with
         the predictions 
         of \protect\cite{Frankfurt}
         for the GRV94(HO) (dashed line) and CTEQ(4M) (full line)
         parametrisations of
         the  proton structure function. The data have
         been scaled taking into account the muonic branching ratio
         and the contributions of the \upsi(2S) and \upsi(3S), 
         as described in the text.
         The inner error bars show the statistical errors, the outer bars
         correspond to the statistical and systematic errors added 
         in quadrature.
         }
\label{fig_results}
\end{center}
\end{figure}

In Figure \ref{fig_results}(a),(b) these values are  compared
with the theoretical calculations  which give $\approx 60$ pb
for $\sigma_{\gamma p \rightarrow \Upsilon(1S)~ p}$
and $\approx 0.001$ for the ratio, both weakly dependent
on the structure function parametrisation used.
Our measurement is higher than the predictions.

%The $\Upsilon(1S)$, $\Upsilon(2S)$
%and $\Upsilon(3S)$ states are produced in different hadronic
%reactions in a range of energies in proportions of $\sigma \cdot \BR$ 
%that are constant within errors \cite{i1}. 

%Under the assumption 
%in agreement with the prediction of reference \cite{Gavai}, 
%detected. Using $ \BR = (2.48 ~\pm ~0.07)$\% for the $\Upsilon(1S)$ \cite{PDG},
%the $\Upsilon(1S)$  photoproduction cross section is 
%$(450~\pm~170~^{+60}_{-80})~\rm pb$ at a mean photon-proton
%center of mass energy of $120$ GeV. 
%The ratio 
%The prediction of the QCD-based model of reference \cite{Frankfurt} 
%is lower at the same energy, $\approx 60 ~\rm pb$, independent of
%the structure function parametrisation used.
%Our measurement is roughly 2 $\sigma$ higher than the predictions
%The model of reference \cite{Frankfurt}, which gives a satisfactory
%description of the \jpsi~data, predicts $\approx 0.001$ for the ratio. 
%In Fig. \ref{fig_results}(a),(b)  the data are  compared
%with the predictions. 
%The data are above the predictions of the available QCD-based models.

\vskip 1.0cm 
%%%%%%%%%%%%%%%%%%%%%
% Acknowledgements  %
%%%%%%%%%%%%%%%%%%%%%

\noindent {\Large\bf Acknowledgments}
\vskip 0.5cm

We thank the DESY Directorate for their strong support and encouragement.
The experiment was made possible by the inventiveness and the diligent
efforts of the DESY machine group.
The design, construction and installation of the ZEUS detector have
been made possible by the ingenuity and dedicated efforts of many
people from inside DESY and from the home institutes who are not
listed as authors.
Their contributions are acknowledged with great appreciation.

We thank L. Frankfurt and M. Strikman for valuable discussions and
R. Engel and M. McDermott for making their tables of theoretical predictions 
available. 

\newpage
%
%--------- REFERENCES -------------
%

\newpage

%\hspace*{-5.cm}
%\begin{table}[P]
\begin{table}
%\begin{minipage}[b]{\textwidth}
\begin{center}
\begin{tabular}{|c|c|c|c|c|c|c|c|} 
\hline 
 &  &  &  &  &  &  & \\ 
$W$ range   & $\langle W \rangle$ &  $N_{evt}$   & \Acce  &
$\sigma_{ep \rightarrow e \Upsilon p} \cdot \BR$ & $\Phi_T$ & 
$\sigma_{\gamma p \rightarrow \Upsilon p}\cdot \BR$ &
$\frac{\sigma_{\gamma p \rightarrow \Upsilon p}\cdot \BR}
     {\sigma_{\gamma p \rightarrow J/\psi p}} \times 10^{4}$\\
\small (GeV) & \small (GeV)  &     &        & \small (pb) &  & \small (pb) &\\   
          &            &           &        &  & &      & \\
\hline
          &            &           &        &  & &      & \\  
\small
$80-160$  & 
\small
$120$      & 
\small
$17.1\pm~7.5$  &
\small
0.43  &  
\small
\sigmabrept  &  
\small
0.051        & 
\small
\sigmabrgpt & 
\small
\ratioupspsito\\ 
          &            &                &        &  & &      & \\

\hline
\end{tabular}
\begin{minipage}[b]{16.8cm}
    \caption{
             The results for the unresolved $\Upsilon$
             cross sections times the muonic branching ratios, 
             $\sum_i \sigma_i \cdot \BR_i$, where $i$ runs over the three states
             $\Upsilon (1S),\Upsilon (2S),\Upsilon (3S)$ and $\BR_i$ is the 
             branching ratio of each state to muons.
             $N_{evt}$ is the number of events after subtraction of the
             non-resonant background contribution: $75$\% of $N_{evt}$
             is attributed to the elastic reaction, the remaining
             $25$\% to the proton dissociative process.
             \Acce~is the acceptance and $\Phi_T$ is 
             the photon flux used to calculate the $\gamma p$ cross
             section $\sigma_{\gamma p \rightarrow \Upsilon p}$ from the $ep$
             cross section $\sigma_{ep \rightarrow e \Upsilon p}$.
             The last column contains the ratio of unresolved $\Upsilon$
             cross section times the muonic branching fraction 
	     to $J/\psi$ cross section.
             The first uncertainties are statistical and the second ones
             systematic. }    
\label{Tab:xsect}
\end{minipage}
\end{center}
%\end{minipage}
\end{table}    

\newpage

\newpage

\newpage

\vfill\eject

\begin{thebibliography}{99}
    
\bibitem{ZeusJPsi94}
    ZEUS Collab., J. Breitweg et al., Z.~Phys.~{\bf C75} (1997) 215.
%
\bibitem{H1JPsi94}
    H1 Collab., S. Aid et al., Nucl.~Phys.~{\bf B472} (1996) 3;\\
    H1 Collab., S. Aid et al., Nucl.~Phys.~{\bf B468} (1996) 3.
%
%\bibitem{Landshoff} A. Donnachie and P.V. Landshoff, Phys.~Lett.~{\bf B296}
%            (1992) 227.
%
\bibitem{Brodsky} S.J. Brodsky et al., Phys.~Rev.~{\bf D50} (1994) 3134.
%
%\bibitem{Collins} J.C. Collins, L. Frankfurt and M. Strikman, Phys.~Rev.~{\bf D56} (1997) 2982.
%
\bibitem{Ryskin} M.G. Ryskin, Z. Phys. {\bf C57} (1993) 89;\\
        M.G. Ryskin, R.G. Roberts, A.D. Martin and E.M. Levin, 
        Z. Phys. {\bf C76} (1997) 231. 
%             
\bibitem{Frankfurt} L. Frankfurt, W. Koepf and M. Strikman, 
         Phys. ~Rev. ~{\bf D57} (1998) 512;\\
         R. Engel and M. McDermott, private communication.
%	 
\bibitem{i2} BCDMS Collab., D. Bollini et al., Nucl.~ Phys.~{\bf B199} (1982) 27;\\
             EMC Collab., J.J. Aubert et al., Nucl.~Phys.~{\bf B213} (1983) 1.
%
\bibitem{CDFups95}
    CDF Collab., F. Abe et al., Phys.~Rev.~Lett.~{\bf 75} (1995) 4358.
\label{CDFups95}
%
\bibitem{i1} J.H. Cobb et al., Phys.~Lett.~{\bf B72} (1977) 273;\\ 
             K. Ueno et al., Phys.~Rev.~Lett.~{\bf 42} (1979) 486;\\
             J. Badier et al., Phys.~Lett.~{\bf B86} (1979) 98;\\
             C. Kourkoumelis et al., Phys.~Lett.~{\bf B91} (1980) 481;\\
             NA10 Collab., S. Falciano et al., Phys.~Lett.~{\bf B158} (1985) 92;\\
             NA10 Collab., M. Grossmann-Handschin et al., Phys.~Lett.~{\bf B179} (1986) 170;\\
             E605 Collab., T. Yoshida et al., Phys.~Rev.~{\bf D39} (1989) 3516;\\
             G. Moreno et al., Phys.~Rev.~{\bf D43} (1991) 2815;\\
             E771 Collab., T. Alexopoulos et al., Phys.~Lett.~{\bf B374} (1996) 271.
%
%\bibitem{i0} S.W. Herb et al., Phys.~Rev.~Lett.~{\bf 39} (1977) 252.
%               
\bibitem{sigtot_photoprod}  ZEUS Collab., M. Derrick et al., Phys.~Lett.~{\bf
B293} (1992) 465;\\ 
                            ZEUS Collab., M. Derrick et al., Z.~Phys.~{\bf C63}
                            (1994) 391.
%
\bibitem{Detector}  ZEUS Collab., The ZEUS Detector, Status Report 1993, DESY
1993.
%
\bibitem{CTD}  N. Harnew et al., Nucl.~Inst.~Meth.~{\bf A279} (1989) 290;\\ 
               B. Foster et al., Nucl.~Phys.~{\bf B}~(Proc.~Suppl.)~{\bf 32} (1993)
               181;\\
               B. Foster et al., Nucl.~Inst.~Meth.~{\bf A338} (1994) 254.
%
\bibitem{CAL}  M. Derrick et al., Nucl.~Inst.~Meth.~{\bf A309} (1991) 77;\\
               A. Andresen et al., Nucl.~Inst.~Meth.~{\bf A309} (1991) 101;\\
               A. Bernstein et al., Nucl.~Inst.~Meth.~{\bf A336} (1993) 23.                      
%             
\bibitem{BRMU} 
               G. Abbiendi et al., Nucl. Inst. Meth.~{\bf A333} (1993) 342.
%
\bibitem{DIPSI} M. Arneodo, L. Lamberti and M.G. Ryskin, 
                    Comp. Phys. Comm.~{\bf 100} (1997) 195.
%                        
\bibitem{Kasprzak} M. Kasprzak, PhD Thesis, Warsaw University, 
                   DESY F35D-96-16 (1994).
%
\bibitem{diffvm} B. List, Diploma Thesis, Techn. Univ. Berlin (1993) unpublished.		   
%
\bibitem{Vermaseren} J.A.M. Vermaseren, Nucl. Phys., {\bf B229} (1983) 347;\\
          S.P. Baranov, O. D\"unger, H. Shooshtari and J.A.M. Vermaseren,
          Proc. of the Workshop `Physics at HERA', Vol. III, 
          Oct. 1991, 1478.
%
\bibitem{geant}   GEANT 3.13: R.~Brun et al., CERN DD/EE--84--1 (1987).
%
\bibitem{PDG}
Review of Particle Properties, Particle Data Group, R.M. Barnett et al., 
Phys.~ Rev.~{\bf D54} (1996) 1. 
%
\bibitem{Flux} 
V.N. Gribov, V.A. Kolkunov, L.B. Okun and V.M. Shekhter, Sov. Phys. ~JETP~ {\bf14} (1962) 1308;\\
V.M. Budnev, I.F. Ginzburg, G.V. Meledin and V.G. Serbo, Phys. Rep. ~{\bf C15} (1975) 181. 
%
%\bibitem{Gavai} R. Gavai et al., Int.~J.~Mod.~Phys.~{\bf A10} (1995) 3043.               
%
\end{thebibliography}
\end{document}